\documentclass[conference]{IEEEtran}
\IEEEoverridecommandlockouts

\usepackage{comment}
\usepackage{booktabs} 
\usepackage{hyperref}
\usepackage{url}
\usepackage{fancyhdr}
\usepackage{comment}
\usepackage{graphicx}
\usepackage{caption}
\usepackage{multirow}
\usepackage[scientific-notation=true, round-precision=2, round-mode=figures]{siunitx}
\usepackage{xspace}
\newcommand{\eg}{e.g.\xspace}
\usepackage{array}
\usepackage{color}
\usepackage{amsmath}
\usepackage{mathtools}
\usepackage[ruled,vlined]{algorithm2e}
\usepackage{subfigure}

\usepackage[shortlabels]{enumitem}
\usepackage{booktabs}
\usepackage{wasysym}
\usepackage{multirow}

\usepackage{siunitx}



\newcommand{\sys}{\mbox{\textsc{CephMaestro}}\xspace}

\def\BibTeX{{\rm B\kern-.05em{\sc i\kern-.025em b}\kern-.08em
    T\kern-.1667em\lower.7ex\hbox{E}\kern-.125emX}}

\usepackage{eso-pic}
\AddToShipoutPictureBG*{%
  \AtPageLowerLeft{%
    \put(0,22){\makebox[\paperwidth]{\parbox{0.86\paperwidth}{\centering\scriptsize
      \copyright~2026 IEEE. Personal use of this material is permitted. Permission from IEEE must be obtained for all other uses, in any current or future media, including reprinting/republishing this material for advertising or promotional purposes, creating new collective works, for resale or redistribution to servers or lists, or reuse of any copyrighted component of this work in other works.\\
      Accepted to the 32nd IEEE International Conference on Parallel and Distributed Systems (ICPADS 2026).}}}}}
      
\begin{document}

\title{\sys: A Safety-Gated AI Agent for Natural-Language Ceph Storage Management
}

\author{\IEEEauthorblockN{1\textsuperscript{st} Vladislav Esaulov}
\IEEEauthorblockA{\textit{Georgia State University} \\
Atlanta, GA \\
vesaulov1@student.gsu.edu}
\and
\IEEEauthorblockN{2\textsuperscript{nd} Lipeng Wan}
\IEEEauthorblockA{\textit{Georgia State University} \\
Atlanta, GA  \\
lwan@gsu.edu}
}

\maketitle

\begin{abstract}
Managing Ceph storage clusters is challenging because operators must reason across interacting subsystems such as OSDs, placement groups, pools, CRUSH topology, monitors, managers, recovery behavior, capacity usage, and performance symptoms. Many tasks require multi-step diagnosis, state-dependent interpretation, and careful verification. Recent advances in large language models create an opportunity to reduce this burden by enabling natural-language Ceph management. However, applying LLMs to Ceph raises two practical challenges. First, storage management is safety-critical: incorrect actions may take services offline, compromise availability or durability, or cause data loss; even less destructive mistakes can trigger unnecessary data movement, recovery storms, or severe client-I/O disruption. Second, large proprietary or high-parameter LLMs can be expensive and slow for interactive system management, motivating designs that work with smaller open-source models.

This paper presents \sys, a safety-gated AI agent for natural-language Ceph storage management. \sys uses an LLM to interpret operator intent, select tools, and construct workflows, while separating model-driven planning from deterministic execution control. The system exposes 112 typed Ceph tools, combines a fast path for simple requests with a bounded ReAct-style loop for multi-step workflows, and mediates state-changing actions through a deterministic action engine that performs risk classification, confirmation gating, dry-run handling, rate limiting, and audit logging. We evaluate \sys using five locally served open-source models and 122 Ceph management prompts covering health inspection, OSD management, pool operations, placement-group diagnosis, CRUSH-related tasks, recovery tuning, and safety-sensitive operations. ReAct-style execution consistently improves workflow completion, with the best model reaching 91.3$\pm$0.5\% goal achievement. A shortcut-ablation study shows that multi-step reasoning improves goal achievement from 65.8\% to 80.2\% even without deterministic shortcuts, while shortcuts further raise ReAct performance to 91.9\% and reduce average latency from 90.3s to 3.8s. \sys's deterministic safety layer achieves 100\% risk-classification accuracy on 95 labeled actions and gates all 37 high-risk workflow cases with zero observed unsafe leaks. These results show that smaller open-source LLMs can support useful Ceph management assistance when model planning is constrained by storage-aware execution control.
\end{abstract}

\begin{IEEEkeywords}
Ceph, storage system management, large language models, AI agents,
tool-augmented reasoning, AI safety
\end{IEEEkeywords}
\section{Introduction}
\label{sec:intro}

Distributed storage systems are foundational infrastructure for cloud platforms, enterprise services, and data-intensive scientific computing. Among them, Ceph~\cite{weil2006ceph} is widely used because it provides scalable object, block, and file storage over commodity hardware. This flexibility, however, creates substantial management complexity. A Ceph cluster exposes many interacting subsystems, including object storage devices (OSDs), placement groups (PGs), pools, CRUSH topology~\cite{weil2006crush}, monitors, managers, metadata servers, recovery mechanisms, device health information, capacity state, and performance counters. Operators must interpret these signals to diagnose failures, plan maintenance, tune recovery, manage capacity, and preserve availability.

Ceph management is difficult because many tasks are not isolated command invocations. Diagnosing degraded PGs may require checking cluster health, mapping PGs to acting OSDs, inspecting OSD status, reasoning about recovery activity, and verifying that the cluster eventually returns to an \texttt{active+clean} state. Replacing an OSD requires determining whether the daemon is down, whether replicas remain available, whether the OSD is safe to destroy, and whether subsequent backfill may overload the cluster. Even when individual Ceph commands are well documented, operators must know which commands to run, in what order, how to interpret their outputs, and when it is safe to proceed.

Recent advances in large language models (LLMs) create an opportunity to reduce this operational burden. An LLM-based assistant can allow operators to express management goals in natural language, such as ``why is my cluster degraded?'', ``is it safe to remove this OSD?'', or ``show me the likely cause of slow operations''. Instead of manually assembling command sequences, the assistant can translate the request into a workflow, gather relevant cluster state, call appropriate tools, and summarize observations in an operator-friendly form. In this paper, we focus on locally served open-source LLMs with relatively small parameter counts because large proprietary or high-parameter models can be expensive, slow, and difficult to deploy inside storage-management environments.

However, applying LLMs to Ceph management raises a safety challenge that is more severe than in ordinary question answering or log summarization. A hallucinated command, omitted precondition, or misclassified intent may take storage services offline, compromise data availability or durability, or cause data loss. Even less destructive mistakes can trigger unnecessary data movement, recovery storms, or severe client-I/O disruption. Therefore, the central question is not simply whether an LLM can generate plausible Ceph commands, but how to design an AI agent that can assist operators with complex Ceph workflows while preventing the model from directly controlling safety-critical storage operations.

\sys addresses this challenge with a safety-gated architecture for natural-language Ceph storage management. The system treats the LLM as a planner rather than an execution authority. The model interprets the operator request, selects tools, and constructs a management workflow, while concrete actions pass through deterministic system components before they can affect the cluster. \sys exposes a finite registry of typed Ceph tools instead of allowing arbitrary shell-command generation. It also uses a deterministic action engine to classify risk, enforce confirmation requirements, support dry-run behavior, apply rate limits, and record audit traces.

\sys further recognizes that different Ceph requests require different execution modes. Simple informational requests, such as checking cluster health or listing OSDs, should be handled quickly. More complex tasks, such as diagnosing degraded PGs or planning maintenance, require iterative observation and reasoning. \sys therefore combines a fast single-tool path with a bounded ReAct-style loop, both mediated by the same typed tool registry and safety gate.

This paper makes the following contributions:
\begin{itemize}
    \item We present \sys, a natural-language AI agent for Ceph storage management that maps operator requests to structured workflows over 112 typed Ceph tools.
    \item We introduce a safety-gated execution architecture that separates LLM-based planning from deterministic execution control, preventing the model from directly executing unsafe storage operations.
    \item We evaluate \sys using five locally served open-source models and 122 Ceph management prompts, showing that ReAct-style execution consistently improves workflow goal achievement over a simple path.
    \item We show that deterministic safety gating provides model-independent protection: the action engine achieves 100\% risk-classification accuracy on 95 labeled actions and gates all 37 high-risk workflow cases with zero observed unsafe leaks.
\end{itemize}

Our results show that smaller open-source LLMs can support useful Ceph management assistance, but only when they are integrated into guarded control frameworks rather than used as unconstrained command generators.

\section{Motivation and Design Goals}
\label{sec:motivation}

The main design question for \sys is how much authority an LLM should have in a Ceph management workflow. A simple approach would place the model directly between the operator and the cluster: the user issues a natural-language request, the model generates Ceph commands, and the system executes them. Although simple, this design is unsafe for storage management. Ceph commands vary widely in risk: some only inspect cluster state, while others change placement, trigger recovery, remove daemons, or delete data. User intent may also be ambiguous; for example, ``fix the degraded
cluster'' could mean diagnose the cause, restart a failed daemon, mark an OSD out, or simply wait for recovery. Prior work has made individual Ceph subsystems programmable---for example, Mantle~\cite{sevilla2015mantle} lets operators inject custom metadata load-balancing policies---but such efforts target one subsystem rather than a unified natural-language management interface. \sys therefore introduces a structured boundary between model-generated intent and executable Ceph actions.

\subsection{Failure Modes of Direct LLM Control}

We identify five failure modes that guide the design of \sys.

\textbf{F1: Command hallucination.}
An unconstrained model may generate invalid Ceph commands, unsupported flags, or shell pipelines outside the intended management interface.

\textbf{F2: Missing preconditions.}
Many Ceph operations require checks before execution. For example, removing an OSD should be preceded by inspecting OSD state, PG state, redundancy, and whether the OSD is safe to destroy.

\textbf{F3: Confusing diagnosis with remediation.}
Action-oriented requests such as ``fix this warning'' may require diagnosis first. A safe assistant should not assume that the user has authorized a mutating remediation action.

\textbf{F4: Unsafe parameter choices.}
The risk of a command often depends on its parameters. Pool settings, recovery tunables, and OSD actions must be inspected before execution.

\textbf{F5: Unbounded agent behavior.}
A multi-step agent may repeatedly call tools or continue probing without progress. Interactive management requires bounded and auditable reasoning.

These failure modes suggest that the LLM should act as a planner within a constrained management framework, not as a Ceph administrator with direct execution authority.

\subsection{Design Goals} 

These failure modes lead to five design goals.

\textbf{G1: Natural-language intent mapping.}
Operators should express Ceph management intent in natural language, and the system should map requests to appropriate inspection or management tools while distinguishing diagnosis from authorized remediation.

\textbf{G2: Bounded tool-mediated interaction.}
The model should not generate arbitrary shell commands. It should interact with Ceph through a finite registry of typed tools with explicit schemas and normalized outputs, giving the framework structured actions to inspect and audit.

\textbf{G3: Support for simple and multi-step workflows.}
Routine status queries should be handled efficiently, while diagnostic workflows should support bounded multi-step reasoning with iteration limits, time budgets, duplicate-call suppression, and stall detection.

\textbf{G4: Planning-execution separation.}
The LLM should interpret requests, select tools, and construct workflows, but concrete actions, especially mutating ones, must pass through deterministic framework logic before reaching Ceph.

\textbf{G5: Conservative safety gating.}
Risky actions should be classified, gated, and logged before execution using confirmation gates, dry-run behavior, rate limits, allow/block policies, and audit records. Unnecessary confirmation is acceptable; unsafe ungated mutation is not.

Together, these goals define \sys as a guarded Ceph management assistant: the LLM improves usability and workflow construction, while deterministic components retain control over executable storage actions.

\section{System Design}
\label{sec:design}

\begin{figure}[!t]
\centering
\includegraphics[width=\columnwidth]{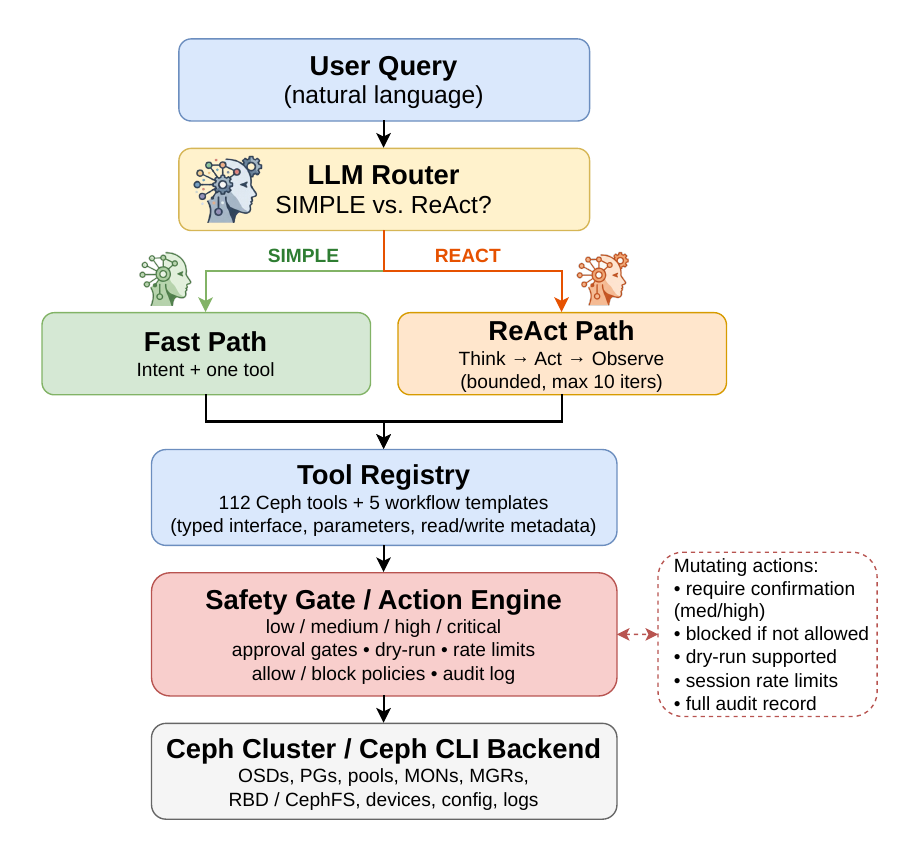}
\caption{\sys architecture. A router selects either a fast single-tool path
or a bounded ReAct path. Both paths use the same typed Ceph tool registry,
and state-changing actions are mediated by a deterministic safety layer
before execution.}
\vspace{-0.4cm}
\label{fig:arch}
\end{figure}

\autoref{fig:arch} shows the \sys architecture. Following the principle in~\autoref{sec:motivation}, the LLM interprets intent, selects tools, and helps construct workflows, but does not directly execute
storage-management commands; state-changing actions are mediated by a deterministic safety engine before they can affect the cluster. We detail each component below.

\subsection{Routing and Workflow Execution}

Ceph requests range from simple status queries to multi-step diagnostics, so \sys uses two execution paths.

\textbf{Simple path.}
Routine requests such as ``list pools'' or ``show OSD status'' are handled through a lightweight routing decision and a single tool-selection step. The selected tool executes once, and the result is formatted for the operator. This path targets read-heavy, low-risk requests that naturally map to one Ceph inspection tool.

\textbf{ReAct path.}
Requests requiring diagnosis, planning, or safety-sensitive reasoning enter a bounded ReAct loop. At each iteration, the model receives the tool schema, the user query, and prior observations. It emits either a \emph{Thought}, an \emph{Action} with JSON-formatted \emph{Action Input}, or a final answer. This lets the model gather state from multiple Ceph subsystems before responding. The loop is bounded by an iteration limit and a session time budget, deduplicates repeated calls, and terminates on stalls.

Both paths are operator-in-the-loop. Read-only requests return concise status information. Diagnostic requests expose selected tools and observations so the operator can audit the result. Mutating requests stop at the confirmation boundary unless dry-run mode or controlled evaluation auto-confirm mode is enabled. Routing affects efficiency and workflow quality, but it does not bypass safety enforcement.

\subsection{Typed Tool Interface and Ceph Backend}

The tool registry is the boundary between model reasoning and cluster control. It exposes Ceph functionality through tool names, descriptions, parameter schemas, and normalized outputs. The prototype registers 112 callable tools spanning cluster health, OSDs, pools, PGs, CRUSH topology, monitors, managers, RBD, CephFS, device health, crash reports, configuration, balancing, diagnostic checks, and workflow-template operations. Tools are implemented through a Ceph cluster manager that invokes standard Ceph commands and normalizes results into Python objects or dictionaries.

The LLM can choose a registered tool and propose parameters, but it cannot construct arbitrary shell commands. This finite interface reduces command hallucination, gives the safety layer stable action names and parameters to inspect, and makes tool use auditable outside the LLM transcript. Higher-level diagnostic helpers and workflow templates are exposed through the same interface; they provide structured guidance for common cases such as down-OSD recovery, degraded-PG investigation, capacity expansion, and performance issues, but they do not bypass the safety engine.

\sys follows three framework contracts. The \emph{tool contract} requires each exposed Ceph operation to have a name, schema, normalized result, and risk-classification target. The \emph{execution contract} requires each request to run through either the simple path or bounded ReAct path. The \emph{safety contract} requires every state-changing operation to cross the deterministic action engine before reaching Ceph. These contracts are independent of the model provider.

\subsection{Safety-Gated Execution}

The safety contract is conservative: the LLM is treated as an untrusted planner rather than an execution authority. It may propose an operation and fill in parameters, but state-changing actions cross a separate policy boundary before execution. This boundary enforces operator policy, such as confirmation before destructive operations or limits on mutating actions, and records denied, dry-run, gated, or completed actions outside the LLM transcript.

The prototype assumes that the operator is authenticated to the node running the assistant and that the underlying Ceph CLI tools are trusted. Multi-tenant authorization, role-specific access control, and site-specific policy integration are outside the current prototype. The prototype also forbids free-form shell generation, turning the Ceph command space into a finite action vocabulary that can be audited and risk-classified.

The \texttt{ActionEngine} classifies actions as \textsc{low}, \textsc{medium}, \textsc{high}, or \textsc{critical}. Read-only actions are low risk. Bounded or reversible changes are medium risk. Actions that may affect availability, trigger data movement, or modify important cluster state are high risk. Destructive operations, such as deleting pools or destroying OSDs, are critical.

\begin{table}[t]
\centering
\small
\begin{tabular}{@{}lll@{}}
\toprule
\textbf{Risk} & \textbf{Examples} & \textbf{Policy} \\
\midrule
Low & health, status, stats & auto-approve \\
Medium & scrub, scoped config & log or confirm \\
High & reweight, restart, create & require confirmation \\
Critical & delete, destroy, purge & confirm and audit \\
\bottomrule
\end{tabular}
\caption{Action-risk policy used by the prototype. The policy is deterministic; the LLM does not decide whether an action is safe.}
\vspace{-0.4cm}
\label{tab:risk-policy}
\end{table}

Risk classification uses deterministic table lookup with selected parameter-aware escalation rules. For example, a pool-parameter change that reduces replication size or \texttt{min\_size} to one is treated more conservatively than an ordinary parameter update. The prototype supports confirmation gates, dry-run mode, per-session rate limits, allow lists, block lists, rollback hints for selected reversible operations, and structured audit records. Each managed action records the original request, selected tool, parameters, risk level, confirmation status, execution outcome, and timestamp.

Thus, \sys reduces command-selection and state-correlation burden without removing the operator from safety-sensitive decisions. Model-generated plans remain useful for workflow construction, but executable storage actions remain under deterministic system control.

\section{Implementation}
\label{sec:implementation}

\sys is implemented in roughly 14{,}000 lines of Python. The code is organized into three layers: an agent layer that implements routing, the ReAct loop, the tool registry, and the action engine; a core layer that interacts with Ceph; and a thin service layer that wires entry points together. All components share a single \texttt{config.yaml} and an \texttt{LLMProvider} abstraction, allowing the model backend and safety policy to change without modifying the Ceph-facing control surface.

\subsection{Model Serving}

\sys treats the LLM as a replaceable component behind a provider interface. The current implementation supports local SGLang~\cite{zheng2024sglang}, OpenAI-compatible APIs, and Ollama. All experiments in this paper use locally served open-source models through SGLang, which exposes an OpenAI-compatible \texttt{/v1/chat/completions} endpoint. We use this setup because large proprietary or high-parameter models can be expensive, slow, and difficult to deploy inside storage-management environments.

Each evaluated model runs on a single NVIDIA A100 GPU with 40\,GB of memory and a 16K-token context window. We use temperature 0.1 and a 2{,}000-token answer budget to reduce variation in tool selection. Each model is referenced by an alias that fixes its weights, quantization, and backend-specific options. This allows the same \sys implementation to evaluate Llama, Qwen, DeepSeek, and related open-source models without changing the Ceph tool layer or safety policy.

\subsection{Tool Registry and Ceph Backend}

The tool registry defines the finite action vocabulary exposed to the model. The current prototype includes 112 typed tools, each represented as an OpenAI-style function schema with a name, description, and typed parameter object. The model can invoke only these registered tools and never emits free-form shell commands.

Each tool binds to a method in a \texttt{ClusterManager}, which invokes the corresponding \texttt{ceph}, \texttt{rados}, or \texttt{rbd} command and normalizes the result into a Python object or dictionary. This backend centralization gives the rest of the framework a stable \texttt{(action\_name, parameters)} representation for every model-selected operation. It also hides differences in CLI output formats from the ReAct loop and gives the safety layer structured actions to inspect.

\subsection{Routing and Bounded ReAct Loop}

\sys first routes each request to either the simple path or the ReAct path. Routing is implemented as a short LLM classification call. Requests classified as simple are handled through a single tool-selection step. Low-confidence simple classifications are escalated to the ReAct path to avoid prematurely answering diagnostic requests.

The ReAct loop has a deterministic control structure even though the model output is probabilistic. At each step, the model receives the tool schemas, user query, and a scratchpad of previous \emph{Thought}/\emph{Action}/\emph{Observation} triples. It must emit either an action with a JSON argument object or a \texttt{Final Answer}. Action arguments are parsed defensively, with lightweight JSON repair for common formatting errors.

The loop is bounded by \texttt{MAX\_ITERATIONS}${=}10$ and \texttt{MAX\_SESSION\_TIME}${=}120$ seconds. A stall counter ends the session after two consecutive steps without a valid tool call, and a per-tool repetition guard rejects more than three identical \texttt{(tool, arguments)} calls. When the scratchpad exceeds eight entries, it is summarized to stay within the context budget. If a limit is reached, the system synthesizes the best available answer from collected observations rather than failing silently.

\begin{algorithm}[t]
\small
\DontPrintSemicolon
\KwIn{query $q$, tool registry $\mathcal{T}$, action engine $E$}
\KwOut{answer, audited action log}
$S \leftarrow \emptyset$ \tcp*{scratchpad}
\For{$i \leftarrow 1$ \KwTo \texttt{MAX\_ITERATIONS}}{
  \lIf{$\textit{elapsed} > \texttt{MAX\_SESSION\_TIME}$}{\textbf{break}}
  \lIf{$|S| > 8$}{$S \leftarrow \textsc{Summarize}(S)$}
  $(a, x) \leftarrow \textsc{LLM}(q, \mathcal{T}, S)$
  \tcp*{action and arguments}
  \If{$a = \texttt{FinalAnswer}$}{\Return $x$}
  \lIf{$a$ \textnormal{repeated} $>3\times$ \textnormal{or stalled} $2\times$}{\textbf{break}}
  $(\textit{ok}, r) \leftarrow E.\textsc{Check}(a, x)$
  \tcp*{risk gate}
  \lIf{$\textit{ok}$}{$o \leftarrow E.\textsc{Execute}(a, x)$}
  \lElse{$o \leftarrow r$ \tcp*{denied / dry-run / needs confirmation}}
  $S \leftarrow S \cup \{(a,o)\}$
}
\Return \textsc{Synthesize}($S$)
\caption{Bounded ReAct control loop.}
\label{alg:react}
\end{algorithm}

\subsection{Deterministic Action Engine}

Every mutating action passes through the \texttt{ActionEngine} before reaching Ceph. Risk classification is implemented as a static lookup in an \texttt{ACTION\_RISK\_MAP} keyed by tool name. Unknown actions default to \textsc{high} risk so that omissions fail conservatively.

The static risk level is refined using parameter-aware escalation rules. For example, \texttt{set\_pool\_param} is \textsc{medium} by default, but setting \texttt{size} or \texttt{min\_size} to one escalates to \textsc{critical} because it removes redundancy. Changing \texttt{pg\_num}, \texttt{pgp\_num}, or \texttt{crush\_rule} escalates to \textsc{high} because it may trigger cluster-wide data movement. Similar escalation rules cover wildcard PG repair, enabling selected manager modules, and creating authentication entities with wildcard capabilities.

Given a risk level, the policy determines whether the action can execute. Actions at or below the \texttt{auto\_approve\_level}, which is \textsc{low} by default, run immediately. Higher-risk actions require confirmation. The evaluation harness can provide confirmation through an explicit auto-confirm flag for controlled tests. The engine also supports dry-run mode, a per-session cap of 20 actions, and optional allow/block lists. Each attempt, whether approved, denied, or dry-run, appends a structured audit record containing the timestamp, action, parameters, risk level, outcome, and, when available, a rollback hint for reversible operations.

\subsection{Reproducibility}

All entry points construct \sys through a shared initialization path. The evaluation suite and the live fault-injection harness use the same path as the deployed service, so measured behavior matches runtime behavior. The live harness can inject and restore selected Ceph faults on a physical cluster, enabling controlled smoke tests without changing the agent implementation.

\section{Evaluation}
\label{sec:evaluation}

We evaluate \sys as an early safety-gated framework for natural-language Ceph management. The evaluation asks five questions: (1) whether the router selects the intended execution mode, (2) whether ReAct improves workflow goal achievement over the simple path, (3) how much deterministic workflow shortcuts contribute, (4) how \sys performs on live fault-injection scenarios, and (5) whether the deterministic safety layer gates risky actions independently of the LLM.

\textbf{Setup.}
Experiments run on a three-node Ceph cluster deployed with
\texttt{cephadm} on Jetstream2~\cite{hancock2021jetstream2}, each node contributing one OSD, with a single monitor and manager. 
The cluster runs Ceph Squid (v19) with RBD and CephFS enabled, and
pools use the default \texttt{host} failure domain at size=3 so that replicas span distinct hosts.
One node carries an NVIDIA A100 GPU (40\,GB) on which we
serve the evaluated models through SGLang, so the agent runs co-located with the cluster it manages.
The evaluated models are
Llama-3.1-8B~\cite{grattafiori2024llama3}, Qwen3.5-9B~\cite{yang2025qwen3},
DeepSeek-R1-14B~\cite{deepseek2025r1}, Qwen3.5-27B, and
GLM-4.7-Flash-31B. 
This reflects our focus on smaller open-source models that run locally, which are more practical for interactive storage management than large proprietary models that can be costly, slow, or hard to deploy.

The paper-facing suite contains 217 active unit and scenario cases,
including 95 deterministic safety cases, 12 adversarial safety prompts,
and routing/workflow cases. The end-to-end workflow comparison uses 122
Ceph management prompts, of which 111 have explicit goal-achievement oracles.
These prompts cover health inspection, OSD and PG diagnosis, pool and
capacity management, CRUSH-related reasoning, configuration inspection,
recovery-related tasks, and safety-sensitive operations.

\textbf{Scoring.}
Each workflow prompt is evaluated by routing choice, completion, tools
used, latency, and goal achievement. Goal oracles check whether the
response and tool trace satisfy the operational intent, such as
diagnosing the named fault, invoking the expected Ceph inspection tool,
or gating a risky write. Thus, a response must be operationally useful,
not merely syntactically valid. An answer that completes but uses the
wrong tool, misses the key fault, or fails to gate a risky action is
counted as unsuccessful.

\begin{table*}[t]
\centering
\scriptsize
\setlength{\tabcolsep}{2pt}
\begin{tabular}{@{}lllllll@{}}
\toprule
\textbf{Model} &
\textbf{Route Acc.} &
\textbf{Simple Goal} &
\textbf{ReAct Goal} &
\textbf{Gate F1} &
\textbf{Simple/ReAct Lat. (s)} &
\textbf{P95 Lat. (s)} \\
\midrule
Llama-3.1-8B & $85.8{\pm}1.3\%$ & $73.3{\pm}0.5\%$ & $80.2{\pm}0.9\%$ & $0.820{\pm}0.015$ & $0.431{\pm}0.003$ / $1.216{\pm}0.630$ & $7.208{\pm}0.249$ \\
Qwen3.5-9B & $94.3{\pm}1.4\%$ & $67.3{\pm}2.3\%$ & $91.3{\pm}0.5\%$ & $0.782{\pm}0.024$ & $2.564{\pm}0.065$ / $4.666{\pm}0.111$ & $34.060{\pm}3.631$ \\
DeepSeek-R1-14B & $83.6{\pm}2.5\%$ & $70.9{\pm}1.9\%$ & $85.3{\pm}0.5\%$ & $0.803{\pm}0.020$ & $1.368{\pm}0.062$ / $1.503{\pm}0.254$ & $30.108{\pm}0.812$ \\
Qwen3.5-27B & $90.2{\pm}0.9\%$ & $69.1{\pm}2.1\%$ & $89.5{\pm}0.5\%$ & $0.819{\pm}0.014$ & $5.313{\pm}0.509$ / $6.448{\pm}0.649$ & $95.836{\pm}2.443$ \\
GLM-4.7-Flash-31B & $87.3{\pm}1.8\%$ & $73.0{\pm}2.7\%$ & $91.0{\pm}1.8\%$ & $0.902{\pm}0.015$ & $3.169{\pm}0.413$ / $7.544{\pm}0.023$ & $156.277{\pm}2.649$ \\
\bottomrule
\end{tabular}
\caption{Model results, reported as mean $\pm$ standard deviation over comparable runs. Route accuracy, goal rates, and workflow gating F1 are measured over 122 workflow prompts, with 111 goal-scored prompts. P95 latency is from the seven-operation latency profile repeated five times per run.}
\vspace{-0.4cm}
\label{tab:model-results}
\end{table*}

\subsection{Routing and Workflow Goal Achievement}

\autoref{tab:model-results} summarizes the main workflow results.
Routing accuracy ranges from $83.6{\pm}2.5\%$ to $94.3{\pm}1.4\%$ across
the five models, with Qwen3.5-9B achieving the highest routing accuracy.
Routing improves efficiency by avoiding unnecessary ReAct traces for
simple requests. However, it is not a safety boundary: both execution
paths use the same typed tool registry and deterministic safety layer.

On the 111 goal-scored prompts, ReAct improves goal achievement over the
simple path for every model. The best result is Qwen3.5-9B, which reaches
$91.3{\pm}0.5\%$ ReAct goal achievement. GLM-4.7-Flash-31B achieves a
similar ReAct goal rate, but with substantially higher tail latency.
These results show that multi-step observation and reasoning are useful
for Ceph workflows, but model choice affects both success and
interactivity.

The improvement from ReAct is largest for prompts that require
correlating multiple Ceph views before answering. The simple path is
effective for direct read-only requests, but it often lacks enough
context for diagnostic or safety-sensitive workflows. Remaining ReAct
failures are mostly due to wrong tool selection, missing follow-up
inspection, or insufficient answer synthesis rather than framework
crashes.

\begin{table}[t]
\centering
\small
\setlength{\tabcolsep}{4pt}
\begin{tabular}{@{}llllc@{}}
\toprule
\textbf{Task category} & \textbf{Tasks} & \textbf{Simple Goal} & \textbf{ReAct Goal} & \textbf{$\Delta$} \\
\midrule
OSD \& recovery   &  15 & 69\% & 96\%  & $+27$ pp \\
PG \& data        &  13 & 69\% & 85\%  & $+16$ pp \\
Pool \& capacity  &  21 & 70\% & 90\%  & $+20$ pp \\
Health \& perf.   &  26 & 69\% & 85\%  & $+16$ pp \\
Auth \& config.   &  10 & 40\% & 90\%  & $+50$ pp \\
CRUSH \& topology &   7 & 52\% & 100\% & $+48$ pp \\
Mon/Mgr/FS        &  19 & 79\% & 100\% & $+21$ pp \\
\midrule
\textbf{Overall}  & 111 & \textbf{67\%} & \textbf{91\%} & \textbf{$+24$ pp} \\
\bottomrule
\end{tabular}
\caption{Per-category workflow goal achievement for Qwen3.5-9B, averaged over three runs. ReAct improves over the simple path in every category, with the largest gains on workflows that require multi-step state gathering.}
\label{tab:per-category}
\end{table}

\autoref{tab:per-category} breaks down Qwen3.5-9B results by task
family. ReAct improves goal achievement in every category, but the gain
depends on task structure. Direct status-oriented categories, such as
Mon/Mgr/FS, start from a stronger simple-path baseline. In contrast,
auth/configuration and CRUSH-topology tasks benefit substantially from
multi-step reasoning because a single tool call is rarely sufficient to
gather the necessary evidence.

\subsection{Workflow Shortcuts and Latency}

\begin{table}[t]
\centering
\small
\setlength{\tabcolsep}{3pt}
\begin{tabular}{@{}llllll@{}}
\toprule
\textbf{Mode} & \textbf{Simple} & \textbf{ReAct} & \textbf{ReAct} & \textbf{Steps} & \textbf{Tool} \\
 & \textbf{Goal} & \textbf{Goal} & \textbf{Lat.} & & \textbf{Cov.} \\
\midrule
Shortcuts on  & 66.7\% & 91.9\% & 3.8\,s  & 2.6 & 78\% \\
Shortcuts off & 65.8\% & 80.2\% & 90.3\,s & 3.5 & 63\% \\
\bottomrule
\end{tabular}
\caption{Qwen3.5-9B shortcut ablation over 122 workflow prompts, with 111 goal-scored cases. Shortcuts improve both success and latency, but shortcut-disabled ReAct still outperforms simple execution.}
\vspace{-0.4cm}
\label{tab:shortcut-ablation}
\end{table}

\autoref{tab:shortcut-ablation} separates model reasoning from
deterministic workflow shortcuts. With shortcuts disabled, ReAct still
improves goal achievement over the simple path by 14.4 percentage points
(80.2\% vs.\ 65.8\%). This shows that the ReAct loop itself contributes
useful multi-step tool use. With shortcuts enabled, ReAct reaches 91.9\%
goal achievement and average latency drops from 90.3\,s to 3.8\,s. The
shortcuts therefore improve both reliability and interactivity by
capturing common Ceph workflow structure in deterministic system logic.

Latency varies significantly by model. Llama-3.1-8B is the most
interactive but has lower ReAct goal achievement than Qwen3.5-9B.
Qwen3.5-27B and GLM-4.7-Flash-31B achieve strong goal rates but have
higher tail latency. This reinforces a practical point for storage
management: larger models are not always preferable, because operators
need timely responses and reliable tool use, not only higher model
capacity.

\subsection{Live Fault-Injection Scenarios}

We also evaluate \sys on live fault-injection scenarios against a
physical Ceph cluster. Each scenario follows the same protocol: inject a
fault, issue a natural-language diagnostic prompt, verify the fault
condition, and restore the cluster. We define 12 scenarios, but skip
\texttt{fi\_mon\_down} because the single-monitor testbed cannot safely
lose quorum, leaving 11 evaluated scenarios. Each model is run five
times, for 55 scenario-runs per model.

A run passes if the agent both invokes at least one expected diagnostic
or remediation tool and produces a response containing expected
fault-specific evidence. This oracle is intentionally stricter than
checking for a plausible natural-language explanation: the agent must
use the cluster-facing tools needed to support the diagnosis.

\begin{table*}[t]
\centering
\scriptsize
\setlength{\tabcolsep}{3pt}
\begin{tabular}{@{}lccccc@{}}
\toprule
\textbf{Scenario} &
  \textbf{Qwen3.5-9B} & \textbf{Qwen3.5-27B} & \textbf{Llama-3.1-8B} &
  \textbf{DeepSeek-R1-14B} & \textbf{GLM-4.7-Flash-31B} \\
\midrule
Stale OSD flags              & 5/5 & 5/5 & 5/5 & 5/5 & 5/5 \\
Pool quota pressure          & 4/5 & 5/5 & 1/5 & 0/5 & 4/5 \\
Quota exhaustion             & 5/5 & 5/5 & 5/5 & 5/5 & 5/5 \\
Risky pool config            & 5/5 & 1/5 & 4/5 & 1/5 & 4/5 \\
Auth caps misconfiguration   & 4/5 & 3/5 & 3/5 & 0/5 & 3/5 \\
Unused CRUSH rule            & 2/5 & 1/5 & 5/5 & 5/5 & 2/5 \\
Manager module disabled      & 5/5 & 5/5 & 5/5 & 5/5 & 5/5 \\
I/O performance load         & 5/5 & 5/5 & 5/5 & 5/5 & 5/5 \\
Stale RBD snapshot           & 4/5 & 3/5 & 2/5 & 0/5 & 4/5 \\
Balancer disabled            & 5/5 & 5/5 & 5/5 & 0/5 & 3/5 \\
OSD failure + recovery       & 5/5 & 5/5 & 5/5 & 5/5 & 5/5 \\
\midrule
\textbf{Overall} & \textbf{49/55} & \textbf{43/55} & \textbf{45/55} & \textbf{31/55} & \textbf{45/55} \\
\textbf{Pass rate}           & \textbf{89\%}  & \textbf{78\%}  & \textbf{82\%}  & \textbf{56\%}  & \textbf{82\%} \\
\bottomrule
\end{tabular}
\caption{Fault-injection pass counts across 11 evaluated live scenarios and five runs per model. \texttt{fi\_mon\_down} is omitted because the single-monitor testbed cannot safely lose quorum.}
\vspace{-0.4cm}
\label{tab:fault-injection}
\end{table*}

\autoref{tab:fault-injection} shows that Qwen3.5-9B achieves the
highest pass rate (89\%), followed by Llama-3.1-8B and
GLM-4.7-Flash-31B (82\%). Five scenarios pass across all models and
runs: stale OSD flags, quota exhaustion, manager-module failure, I/O
performance load, and OSD failure with recovery. These scenarios
represent common operational symptoms where the expected diagnostic
tools are relatively direct.

More difficult cases include pool quota pressure, CephX capability
inspection, stale RBD snapshot detection, and some CRUSH-related
queries. Failures mainly arise from wrong tool selection or insufficient
answer synthesis after retrieving the relevant state. These results show
that smaller open-source models can handle many live Ceph diagnostic
tasks, but tool-use reliability remains model- and scenario-dependent.
They also show that model scale alone is not a reliable predictor:
Qwen3.5-9B outperforms the larger Qwen3.5-27B on the aggregate live
benchmark.

\subsection{Safety Behavior}

The deterministic safety evaluation covers 95 labeled actions across
four risk levels: 25 low, 14 medium, 31 high, and 25 critical. The
\texttt{ActionEngine} achieves 100\% risk-classification accuracy and
98.9\% confirmation-classification accuracy. Results are identical
across models because risk classification is implemented by deterministic
policy logic rather than by the LLM.

In the workflow setting, confirmation gating achieves perfect recall:
all 37 high-risk workflow cases are gated, with zero observed unsafe
leaks. Precision is lower because the agent sometimes triggers
unnecessary gates on benign prompts. This is the intended conservative
failure mode for storage management: an unnecessary confirmation slows
the operator, while a missed gate may allow an unsafe cluster mutation.
Together, these results support the central design choice of \sys:
LLMs can help construct Ceph workflows, but safety-sensitive execution
should be enforced by deterministic storage-aware control logic.

\section{Related Work}
\label{sec:related}

\subsection{Tool-Augmented and Agentic LLMs}

Recent work augments LLMs with external tools and APIs. Toolformer~\cite{schick2023toolformer} studies self-supervised API-use learning, Gorilla~\cite{patil2023gorilla} improves API selection with retrieval over documentation, and ToolLLM~\cite{qin2024toolllm} scales tool use to large API collections. Agentic frameworks further interleave reasoning and acting. ReAct~\cite{yao2023react} uses a thought--action--observation loop for multi-step tasks, while broader agent frameworks provide scheduling, memory, context management, and access-control mechanisms for general-purpose agents~\cite{wang2024agentsurvey,mei2024aios}.

These systems show that LLMs can select tools and reason over observations, but they do not target safety-critical storage management. In general API settings, a tool call is usually a task action; in Ceph, a tool call may restart daemons, change placement, trigger recovery, or delete data. \sys adopts tool-augmented and ReAct-style execution, but constrains the model to a finite registry of typed Ceph tools and places a deterministic, storage-aware safety gate between model-selected actions and cluster execution.

\subsection{Benchmarks for Agents Acting on Systems}

Agent benchmarks increasingly evaluate effects on an environment rather than surface-form answers. SWE-bench~\cite{jimenez2024swebench} checks whether generated patches resolve real GitHub issues. AgentBench~\cite{liu2024agentbench} evaluates agents across interactive environments such as operating systems, databases, and web tasks. $\tau$-bench~\cite{yao2024taubench} compares final database state against annotated goals and measures reliability over repeated trials.

Our evaluation follows this state-aware view: goal oracles check whether the response and tool trace satisfy the Ceph management intent, and live fault-injection scenarios test behavior against an actual cluster. The difference is the cost of failure. In Ceph, an incorrect action may cause data movement, recovery storms, outages, or data loss. Therefore, \sys evaluates task success together with deterministic safety gating, rather than treating goal achievement alone as sufficient.

\subsection{LLMs for Systems and Data Management}

LLMs have also been applied to data-management and systems-operation tasks. LLMDB~\cite{zhou2024llmdb} combines fine-tuning, retrieval, and multi-round inference for query rewriting, diagnosis, and analytics. StorageXTuner~\cite{lin2025storagextuner} uses LLM agents to tune storage engines such as RocksDB, LevelDB, CacheLib, and MySQL InnoDB through sandboxed search and lightweight configuration checkers.

These systems demonstrate the promise of LLMs for infrastructure management, but their scope differs from ours. Prior work mainly targets query assistance, diagnosis, analytics, or configuration tuning. \sys targets cluster-level Ceph management, including OSD state, placement groups, pools, CRUSH topology, recovery behavior, and safety-sensitive administrative actions. Unlike sandboxed tuning, \sys is designed for workflows that may act on a live storage cluster, making prevention of unsafe state changes a first-class design requirement.

\subsection{LLM-Based Root-Cause Analysis for Operations}

Another line of work applies LLMs to incident diagnosis and root-cause analysis in cloud operations. Prior studies show that LLMs can infer likely causes and mitigation steps from incident metadata and operational context~\cite{ahmed2023rca,roy2024rca}. RCAgent~\cite{wang2023rcagent} extends this direction with an autonomous tool-augmented agent for data collection and analysis.

These systems are complementary to \sys. They focus mainly on diagnosis and recommendation, while \sys must also manage the boundary between recommendation and execution. In Ceph, moving from diagnosing an issue to performing a fix may affect durability, availability, and data placement. \sys therefore treats the LLM as an untrusted planner and relies on deterministic execution control for mutating actions.

\subsection{Summary}

Compared with prior tool-augmented agents, systems-management agents, and RCA frameworks, \sys focuses on natural-language management of a stateful distributed storage system. Its main distinction is the combination of bounded workflow construction, a typed Ceph tool registry, and a deterministic storage-aware safety layer. This design allows smaller locally served open-source LLMs to assist with Ceph management while preventing the model from directly controlling unsafe storage operations.

\section{Conclusion}
\label{sec:conclusion}

We presented \sys, a safety-gated AI agent for natural-language Ceph management. \sys treats the LLM as an untrusted planner: it interprets operator requests, selects tools, and constructs workflows over 112 typed Ceph tools, while a deterministic action engine retains execution authority. The engine classifies risk, enforces confirmation and dry-run policies, applies rate limits, and records audit logs before state-changing actions reach the cluster. To balance responsiveness and reasoning depth, \sys combines a fast single-tool path for routine requests with a bounded ReAct loop for diagnostic and multi-step workflows.

Across five locally served open-source models and 122 Ceph management prompts, ReAct consistently improves goal achievement over the simple path, with the best model reaching $91.3{\pm}0.5\%$. Deterministic workflow shortcuts further raise performance to $91.9\%$ while reducing average latency from $90.3$\,s to $3.8$\,s. The safety layer achieves $100\%$ risk-classification accuracy on 95 labeled actions and gates all 37 high-risk workflow cases with zero observed unsafe leaks. Live fault-injection experiments also show that smaller open-source models can handle many practical Ceph diagnostic tasks, although reliability remains model- and scenario-dependent. A 9B model matches or outperforms larger models in several settings, showing that parameter count alone is not a reliable predictor of tool-use quality.

These results support the central argument of this paper: LLMs can reduce the burden of complex storage-management workflows, but safety-sensitive execution should remain under deterministic, storage-aware control. \sys is an early prototype; future work will address prompt injection, multi-tenant authorization, role-specific policies, lower-latency monitoring, and evaluation on larger production clusters.

\section*{Acknowledgment}
This material is based upon work supported by the National Science Foundation under Grant No. \texttt{2531903}.

\bibliographystyle{IEEEtran}
\bibliography{references}

\end{document}